\documentclass{ametsocV6.1}
\pdfoutput=1

\title{Can Transfer Learning be Used to Identify Tropical State-Dependent Bias Relevant to Midlatitude Subseasonal Predictability?}

\authors{Kirsten J. Mayer\aff{a},\correspondingauthor{Kirsten J. Mayer, kjmayer@ucar.edu} 
Katherine Dagon\aff{a},
and Maria J. Molina\aff{b}}

\affiliation{\aff{a}{U.S. National Science Foundation National Center for Atmospheric Research} \\
\aff{b}{University of Maryland, College Park}}

\abstract{Previous research has demonstrated that specific states of the climate system can lead to enhanced subseasonal predictability (i.e., state-dependent predictability). However, biases in Earth system models can affect the representation of these states and their subsequent evolution. Here, we present a machine learning framework to identify state-dependent biases in Earth system models. In particular, we investigate the utility of transfer learning with explainable neural networks to identify tropical state-dependent biases in historical simulations of the Energy Exascale Earth System Model version 2 (E3SMv2) relevant for midlatitude subseasonal predictability. Using a perfect model framework, we find transfer learning may require substantially more data than provided by present-day reanalysis datasets to update neural network weights, imparting a cautionary tale for future transfer learning approaches focused on subseasonal modes of variability.}

\begin{document}

\maketitle


\statement
In this study, we explore how two machine learning methodologies (transfer learning and explainable artificial intelligence) can be used to identify biases in Earth System models. We are particularly interested in identifying biases associated with states in our climate that can lead to enhanced predictability on two week to two month timescales. Identifying and then subsequently correcting these biases can lead to improved prediction skill on these timescales. However, we find that our machine learning framework for identifying biases requires more data than our observational datasets provide. 


\section{Introduction}
Subseasonal timescales, spanning approximately two weeks to two months, provide actionable lead times for various sectors such as energy and water management \citep{CWhite2017, CWhite2021}. While skillful forecasts on these timescales are societally important, subseasonal forecasting remains difficult. To improve subseasonal prediction skill, one area of research has explored modes of variability shown to enhance predictability when present \citep[i.e., state-dependent predictability;][]{Mariotti2020}. These modes include phenomena like the Madden-Julian Oscillation \citep[MJO;][]{Madden1971,Madden1972,Madden1994,Tseng2018} and the El Ni\~{n}o-Southern Oscillation \citep[ENSO;][]{Trenberth1997,Johnson2014,Wang2019,chapman2021monthly}. Both these modes of variability influence the northern hemisphere weather during boreal winter through the development and propagation of Rossby waves \citep[e.g.,][]{philander1985nino,Hoskins1993,henderson2016influence,Mundhenk2018,chapman2021monthly}

Previous work has demonstrated that explainable neural networks can identify these states of enhanced predictability in both Earth system models and observations on a variety of timescales \citep{Mayer2021,Mayer2022,Gordon2021}, allowing us to explore sources of predictability unconstrained by predefined indices. However, neural networks learn best in data-rich settings and often require more observational data than are available for prediction at longer timescales. While Earth system model output can be used as surrogates for observational data, these physics-based models have biases that can affect the representation of modes of variability and their subsequent impacts \citep[e.g.,][]{kang2020role,wei2021tropical}, which can hinder the ability of the neural network to learn physically meaningful relationships.

One way to combat this issue is through transfer learning \citep{Tan2018}. Transfer learning attempts to exploit information obtained from training on a problem with a large amount of data, and utilize the learned information for a similar, data-limited application. In climate science, transfer learning can benefit machine learning applications in settings with limited observational data but plentiful Earth system model datasets. We can obtain information about relationships in the Earth system from the vast amount of climate model simulations, and then fine-tune this learned information using observations or reanalysis datasets, often to obtain better performance \citep[e.g.,][]{Ham2019,Kadow2020,Ham2021}. While performance is important, we are specifically interested in whether transfer learning along with eXplainable Artificial Intelligence (XAI) could be used to identify biases in Earth system models. 

Previous research has demonstrated the utility of XAI for exploring learned relationships across machine learning applications ranging from prediction to detection \citep{mcgovern2019making, Molina2021, Davenport2021, Gordon2021, Rader2022}. Here we use XAI to examine how regions learned as important for a prediction problem when initially trained on climate model output may update when the neural network is fine-tuned with a newly introduced observational-based dataset (e.g., reanalysis). In doing so, we can identify differences or biases in regions important for the prediction problem of interest between the two datasets. 

In this paper, we investigate whether transfer learning and XAI can be used to identify tropical state-dependent bias relevant to midlatitude subseasonal predictability in the Energy Exascale Earth System Model version 2 (E3SMv2) historical simulations \citep{Golaz2022}. Specifically, we are interested in tropical precipitation biases associated with subseasonal predictability of anomalous mid-level tropospheric circulation over the North Pacific. To explore the feasibility of this approach, we utilize a ``perfect model" experimental design. This allows us to create artificial biases to correct during transfer learning and quantify the amount of data needed to identify state-dependent biases in tropical sources of midlatitude subseasonal predictability in climate models. While this methodology proves successful in the perfect model framework, more data is required than what is available in observational products, such as present-day reanalyses, to implement this methodology for subseasonal timescales. These results suggest that transfer learning is not ideal for identifying subseasonal state-dependent biases in Earth system models, providing a cautionary tale for those interested in applying transfer learning on subseasonal or longer timescales using models with large state-dependent biases.

\section{Data \& Methods}
\subsection{Data and Pre-processing}
Tropical precipitation (20°S-20°N) and 500hPa geopotential height (z500) over the North Pacific (30-60°N, 170-240°E) are obtained from the 21 ensemble members of the E3SMv2 historical simulations \citep[1950-2014;][]{Golaz2022}. For simplicity, the members will be referred to as members 1 to 21. However, the list of corresponding E3SMv2 member labels is included in Table A1. Both variables are regridded to 2.5$^\circ$x2.5$^\circ$ using bilinear interpolation and each member is detrended and deseasonalized by subtracting a 3rd-order polynomial fit of the ensemble mean for each day of the year. Subsequently, a 7-day running mean is applied to z500 and tropical precipitation. Both running means are not centered, where the past six days are included in the running mean of precipitation and the future six days are included for z500.

Daily tropical precipitation is further standardized at each grid point by subtracting the daily mean and dividing by the daily standard deviation calculated from the training members (see section 2c for training, validation, and testing split details). For z500 averaged over the North Pacific region, the median of the training data is calculated and subtracted such that there are an equal number of positive and negative anomalies within the training dataset. Since this is a categorical prediction (see section 2c for methods), these anomalies are then converted (i.e., integer encoded) into 0s and 1s for negative and positive anomalies, respectively. For the validation and testing data, the anomalies are randomly subset to an equal number of 0s and 1s, ensuring the random chance for each dataset is 50\% (i.e., balanced classes).

To define an ENSO event, we use the non-detrended Ni\~{n}o 3.4 Index provided by the Climate Variability Diagnostics Package for Large Ensembles \citep[CVDP-LE;][]{Phillips2020} and subset the index for the same years as the historical simulation (1950-2014). We then detrend the data using a 3rd-order polynomial fit to the ensemble mean for each day of the year, as was done for precipitation and z500. The monthly samples are resampled to daily data by assigning all days in a given month the Ni\~{n}o 3.4 index value for that month. Any Ni\~{n}o 3.4 index value greater than +1$^\circ$C is classified as an El Ni\~{n}o (21\% of days on average) and any index values less than -1$^\circ$C are classified as La Ni\~{n}a (22\% of days on average).

\begin{figure}[t]
  \noindent\includegraphics[width=30pc,angle=0]{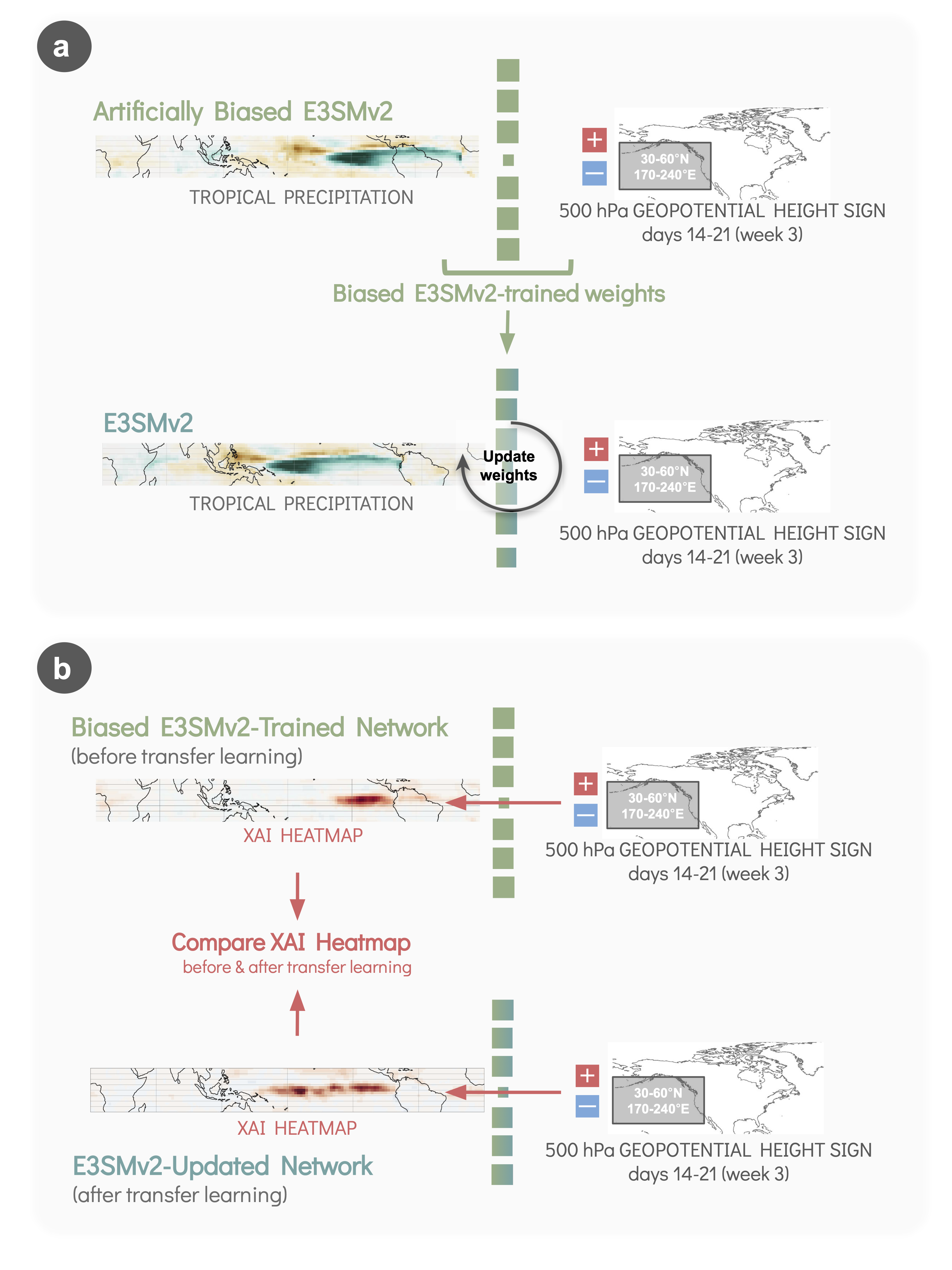}\\
  \caption{Schematic of the method used to identify tropical state-dependent biases relevant for North Pacific subseasonal predictability. Panel (a) details the artificial neural networks pre- and post-transfer learning using the perfect model approach and panel (b) demonstrates the subsequent application of XAI to identify the biases in tropical sources of predictability.}\label{schematic}
\end{figure}

\subsection{Neural Network Architecture}
To explore state-dependent bias for midlatitude subseasonal predictability, we input tropical precipitation into an artificial neural network to predict the sign of the 14-21 day averaged (week 3) atmospheric circulation at 500hPa over the North Pacific. Through a manual hyperparameter search, we select a single-layer artificial neural network with 128 nodes for both the pre- and post-transfer learning neural networks. The rectified linear unit (ReLU) activation function is applied to each node output from the hidden layer. The softmax activation function is applied to the last dense layer, converting the two neural network outputs into values that sum to one, each value representing the likelihood of a negative or positive z500 anomaly. The output value is a measure of the network's confidence in a given prediction, where values closer to one indicate higher confidence. 

For both training and retraining, we use sparse categorical cross-entropy as the loss function, the Adam Optimizer \citep{kingma2014adam} with a learning rate of 0.0001 for gradient descent, and a batch size of 256. To minimize the potential of overfitting, we use a dropout rate of 0.9 and 0.95 for the pre- and post-transfer learning neural networks, respectively, and employ early stopping when the validation loss does not improve over 30 epochs, reverting the previous best validation loss. We also experimented with smaller dropout rates as well as various sized and dynamic learning rates, but found those configurations achieved worse performance (not shown). To test the sensitivity of our neural network to randomly initialized weights, we train 10 neural networks each initialized with different random seeds and find similar performance across seeds. The results presented in this work include all 10 networks.

We note that we obtain similar accuracy on the pre-transfer learning network when using a convolutional neural network. However, in addition to being more computationally expensive to train, the convolutional neural network's performance was worse than the artificial neural network after transfer learning. As a result, we use an artificial neural network where all weights are allowed to update during retraining. A preliminary analysis of this method with a convolution neural network shows similar relevance heatmaps to those from the artificial neural network (Appendix Figure\textbf{ \ref{XAI_EXP3_CNN}}). We leave exploration of the impact of model architecture on these results for future work.

\subsection{Methods}
Standardized weekly mean tropical precipitation anomalies are input into an artificial neural network to predict the sign of the following 14-21 day averaged (week 3) z500 anomaly over the North Pacific during extended boreal winter (November - March). We are interested in tropical precipitation biases that can degrade subseasonal predictability of atmospheric circulation over the North Pacific, where ``bias" is used to describe systematic differences in tropical precipitation between the pre- and post-transfer learning neural networks. 


To explore the feasibility of identifying state-dependent bias with transfer learning and XAI, we use a perfect model framework, where an artificial state-dependent bias is added to tropical precipitation data in half (ten) of the ensemble members. While we find that the accuracy of the pre-transfer learning network stabilizes after about six training members (not shown), using ten members ensures the pre- and post-transfer learning networks have the same number of available training members. We focus on adding bias to tropical precipitation because it is closely associated with the state of ENSO, which is an important driver of subseasonal predictability in the North Pacific on this timescale \citep{kumar1998annual,chapman2021monthly,Mayer2024}. 

Tropical precipitation for the pre-transfer learning neural network are shifted 60$^{o}E$ during an El Ni\~{n}o or La Ni\~{n}a event (Figure \ref{schematic}a, upper network). These shifted precipitation ensemble members will be referred to as the ``biased ensemble members". Nine members are used for training, one member for validation and the remaining member for testing (e.g. training on biased members 1-9, validation on biased member 10, and testing on biased member 21). After initial training with the biased members, the weights from this network are transferred and used to initialize the second network rather than using random weights (Figure \ref{schematic}a, lower network); we refer to this second network as the ``post-transfer learning network''. This second network is subsequently trained further using the non-biased, remaining 10 members: nine members for training, one member for validation, and the remaining member for testing (e.g., retraining on unbiased members 11-19, validation on unbiased member 20, and testing on unbiased member 21).

XAI heatmaps calculated with Integrated Gradients \citep{Sundararajan2017} are produced from the pre-transfer learning network and compared to heatmaps from the post-transfer learning network to identify possible differences in sources of predictability between the biased and unbiased test members (Figure \ref{schematic}b). The Integrated Gradients XAI method is used to create heatmaps, as it has been shown to well represent the relevant contributions for a prediction in climate-like applications \citep{Mamalakis2022}. Each heatmap is normalized by dividing its absolute maximum value before compositing so that the samples are equally represented in the average. 

Previous research has shown that when accuracy increases with confidence, higher network confidence can be used to identify periods of enhanced predictability \citep{Mayer2021}. Therefore, XAI heatmaps for the 20\% most confident and correct neural network predictions are used to explore \textit{state-dependent bias}, following \cite{Mayer2022}. We use confidence rather than compositing on the known biased El Ni\~{n}o and La Ni\~{n}a samples, as biased samples are not necessarily known in real-world applications. However, we find that about 90\% of the confident samples are also El Ni\~{n}o and La Ni\~{n}a days, confirming the utility of this approach for our perfect model application.

To quantify the number of ensemble members (and years) required to detect state-dependent biases, the pre-trained network is updated incrementally, first with one unbiased ensemble member, then two unbiased members, through nine unbiased members, and the skill of each post-transfer learning network on the \textit{unbiased} test member is calculated. The skill of these post-transfer learning networks is then compared to the performance of the pre-trained network on the \textit{biased} testing member. Since the only difference between the biased and unbiased test member is the shifted precipitation during El Ni\~{n}o and La Ni\~{n}a days, the post-transfer learning network should be able to achieve the same accuracy on the unbiased test member as the pre-transfer learning network on the biased test member once the network corrects for the shift. The number of ensemble members needed for successful detection of state-dependent biases is selected when the skill of the pre-transfer learning network on the \textit{biased} testing member is comparable to the post-transfer learning network on the \textit{unbiased} testing member. When this occurs, it suggests the post-transfer learning network has either unlearned the state-dependent bias or has learned to additionally focus on the un-shifted region in the unbiased member. We define the pre- and post-transfer learning model skill as comparable when the 75th percentile of accuracy for the unbiased testing member evaluated on the post-transfer learning model meets or exceeds the accuracy at the 25th percentile of the biased testing member on the pre-transfer learning model. We use the upper and lower quantiles for the definition rather than the median of the accuracy to account for the spread in model skill. We explore the evolution of the XAI heatmaps across the increasing number of retraining ensemble members to determine whether the neural network identifies a new region of importance farther west than the artificially east-shifted region.

For robustness, this analysis is conducted using k-fold cross validation such that each ensemble member is used as the test member once. Only one k-fold example is included in the main text, however, summary plots for the other k-fold test members are included in Figure \ref{performance}c,d, and another test member example is included in the Appendix (Figures \ref{performance_EXP5}, \ref{XAI_EXP5_neg} and \ref{XAI_EXP5_pos}). The results shown here are for training on biased members 1-9, validation on biased member 10, and testing on biased member 21. Retraining uses unbiased members 11-19, validation on unbiased member 20, and testing on unbiased member 21.

\section{Results}

\subsection{Transfer Learning Performance}
The performance of the pre-transfer learning model is first evaluated to confirm it can identify state-dependent predictability. To do so, the model accuracy across confidence levels is calculated. We find that accuracy increases with confidence, suggesting the network can identify states of enhanced predictability \citep[Figure \ref{performance_ann1};][]{Mayer2021}. We also note that the pre-transfer learning model performance on the biased test member is better than the same model's performance on the unbiased version of the same test member (Figure  \ref{performance_ann1}). This result confirms that tropical precipitation associated with ENSO is important for North Pacific subseasonal predictability. Therefore, transfer learning provides an opportunity to improve the performance of the network on unbiased data.  

\begin{figure}[t]
  \noindent\includegraphics[width=39pc,angle=0]{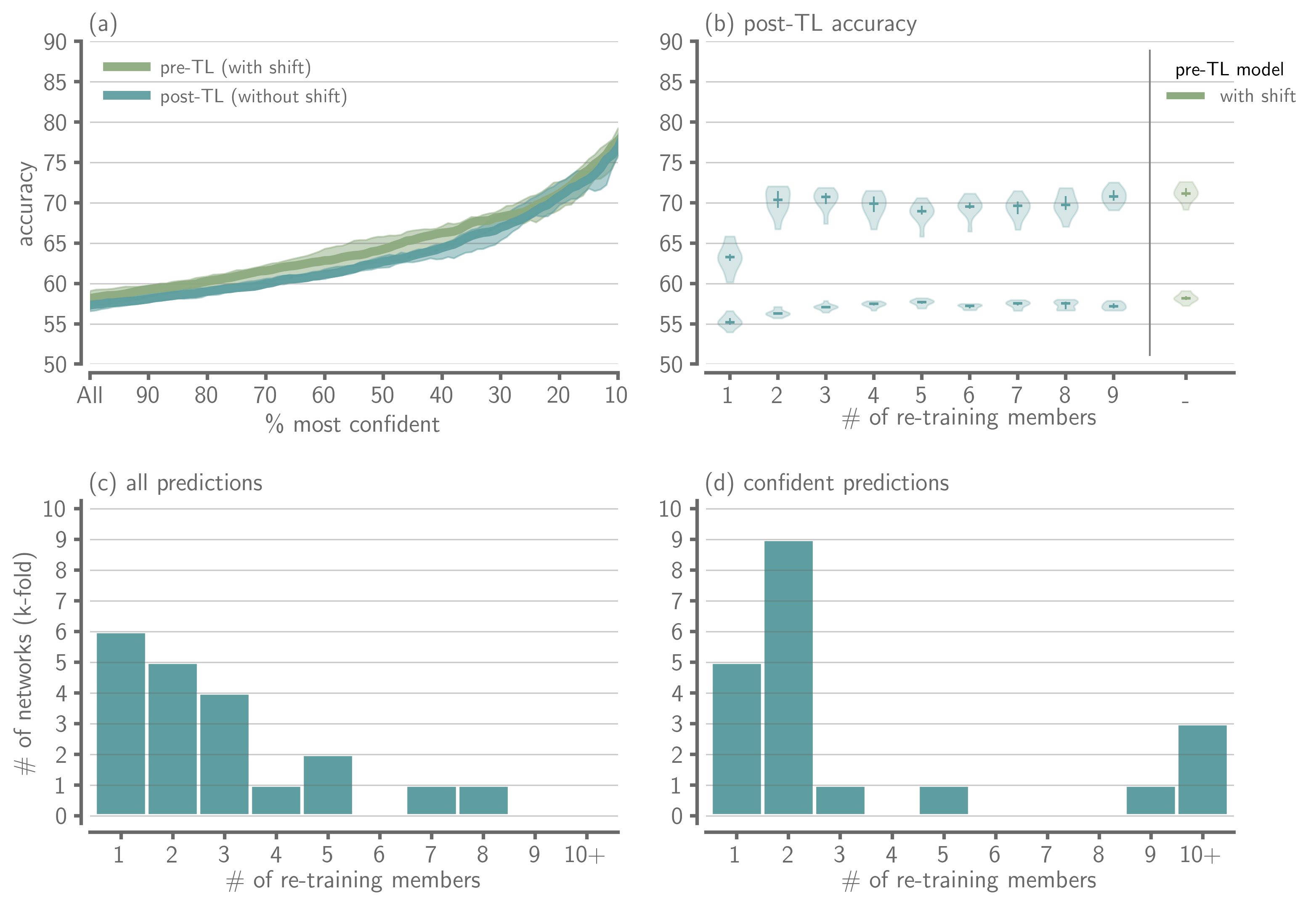}\\
  \caption{(a) Accuracy of the pre-transfer learning (green) and post-transfer learning (teal) network on the bias and unbiased E3SMv2 testing member, respectively, across various confidence thresholds. The thick line represents the median accuracy and the shading represents the spread in accuracy across the 10 seeds. (b) Violin plot of the accuracy across network seeds for all predictions (lower) and the 20\% most confident predictions (upper). The right-most violin plot (green) is the performance of the pre-transfer learning network on the biased test member. The nine teal violin plots are the performance of the post-transfer learning network on the unbiased test data as the number of re-training members increases from one to nine. (c) Count of re-training members needed until the percentile criteria is met [see text for details]. (d) as in (c), but for the 20\% most confident predictions.}\label{performance}
\end{figure}

Transfer learning is next applied using nine, unbiased retraining ensemble members. If transfer learning is successful, we expect the performance of the post-transfer learning network evaluated with the unbiased testing ensemble member to be equal to or better than the performance of the pre-transfer learning network using the same biased test member. Indeed, we find that the performance of the post-transfer learning network using the unbiased test ensemble member has similar accuracy across all confidence levels (Figure \ref{performance}a). Even so, nine training members is equivalent to 576 years of data, which is more than 10 times the length of the observational record from reanalysis data. To analyze the utility of this methodology for real-world applications, we next examine how many retraining members would be required to identify state-dependent biases using transfer learning.

\subsection{How many retraining members are needed?}
To evaluate the number of retraining members required to identify state-dependent biases using transfer learning, we incrementally increase the number of unbiased retraining members from one to nine, calculating the skill on the same unbiased test member each time. The minimum number of members is identified when the 25th percentile of the accuracy of the biased test member on the pre-transfer learning network (bottom of the vertical line in the green violin plot) is equal to or below the 75th percentile of the accuracy of the unbiased test member on the post-transfer learning network (top of the vertical line in the teal violin plots; Figure \ref{performance}b). We find that at least two members (130 years) are required to meet this criteria, using the unbiased test member 21 for confident predictions (top row of violin plots in Figure \ref{performance}b). Expanding this analysis with k-fold cross-validation (k=20, where k is the ensemble member used for testing), we find that most testing members require at least two retraining members to achieve similar accuracy to the pre-transfer learning model for all and the 20\% most confident predictions (Figure \ref{performance}c,d). These results suggest that considerably more observational data is needed before this methodology can be successfully applied to similar problems using reanalysis.

\begin{figure}[t]
  \noindent\includegraphics[width=39pc,angle=0]{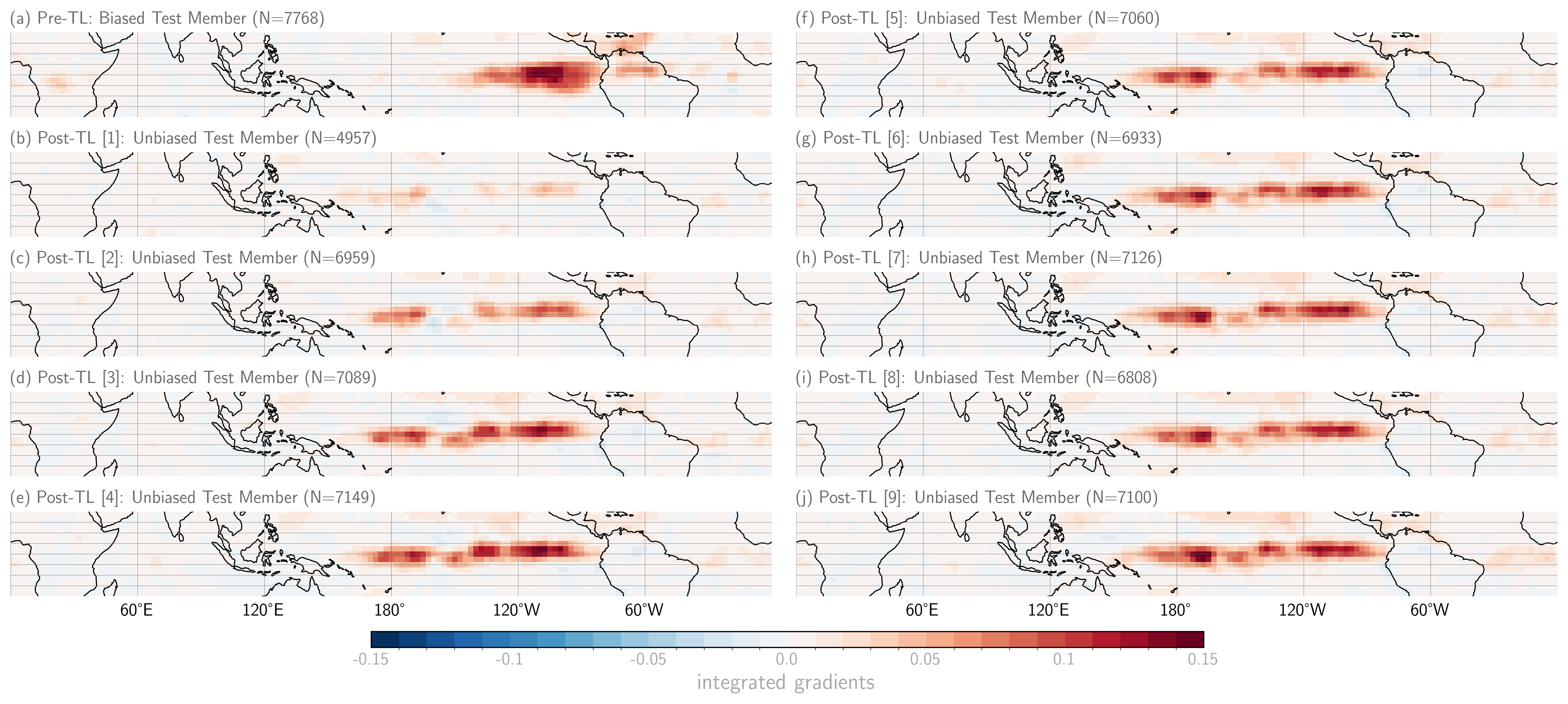}\\
  \caption{(a) Integrated gradients heatmap of the 20\% most confident negative predictions for the pre-transfer learning network on the biased testing member (N=samples in each composite). (b-i) as in (a), but for the unbiased testing member using the post-transfer learning networks retrained with an increasing number of training members (number in brackets). The heatmaps are smoothed with a 2D Gaussian filter with a sigma of 0.75 for ease of interpretation.}\label{XAI}
\end{figure}

While accuracy can be used as a metric to evaluate the performance of transfer learning, it does not explain what information the neural network is using for accurate predictions nor how that information changes as transfer learning progresses. We expect that the relevant regions of the input considered by the post-transfer learning neural network for correct, confident predictions will move westward as the number of unbiased retraining ensemble members increases. To explore the evolution of relevant regions for predictability with increasing retraining members, heatmaps of the integrated gradients for each retraining neural network (averaged across all seeds) are created for the 20\% most confident negative (z500 anomaly) predictions. After one retraining member, a secondary hotspot emerges approximately 60$^{o}W$ of the pre-transfer learning hotspot (180$^{o}$; Figure \ref{XAI}a,b). After two retraining members, this western hotspot relevance is larger and does not evolve further with increasing retraining members after three retraining members (Figure \ref{XAI}b-i). We find that the original eastern hotspot (shown in Figure \ref{XAI}a) seems to disappear after one retraining member and then reemerges throughout the transfer learning iterations. On the other hand, the hot spot over northern South America disappears and does not reemerge, suggesting that it has been ``unlearned" through transfer learning. We also find that positive predictions show similar results, but the western hotspot requires more members to emerge (Figure \ref{XAI_pos}).

To explore whether this eastward extension is a reemergence of the original hotspot learned from biased members or an artifact of training on the unbiased members, an artificial neural network (with the same hyperparameters) is trained with the last nine unbiased members of the E3SMv2 ensemble. We obtain a similar result to Figure \ref{XAI}j (Figure \ref{noTLIG}), suggesting that the region is not a reemergence, but rather learned from the unbiased training data. We hypothesize that the extended eastward signature in the post-transfer learning model is a result of precipitation signals that occur during ENSO-like events, but are not classified as an El Ni\~{n}o or La Ni\~{n}a\ due to our ENSO definition. As a result, the associated precipitation signal is not artificially shifted eastward in the training data for the pre-transfer learning model, and therefore, is also not as strongly present in the pre-transfer learning integrated gradients. This is a limitation of the perfect model approach, and not of the transfer learning methodology. Overall, it appears the transfer learning led to unlearning the original pattern in favor of the new dataset, which may explain why more training data than provided by current-day reanalysis is required for this application of transfer learning.


These results suggest that transfer learning in combination with XAI can be utilized to identify state-dependent subseasonal predictability bias in tropical precipitation. Further, since more than one ensemble member was needed to perform the transfer learning and detect the state-dependent bias with XAI, more data (i.e., samples) are required in an observational application (e.g., using reanalysis) at the subseasonal timescales considered here. We also explore the heatmaps for a testing member which requires more than nine retraining members to meet the accuracy criteria (Figure \ref{XAI_EXP5_neg}) and find the heatmaps look very similar to Figure \ref{XAI}, even though the accuracy is lower. This suggests accuracy \textit{and} XAI heatmaps should be used when identifying whether this method is successful.

\section{Conclusions}
In this paper, we present a methodology to identify state-dependent midlatitude subseasonal predictability biases within climate models, which can be successfully applied with enough data. Through the application of explainable neural networks and network confidence, we can identify these biases for state-dependent predictability without constraints from predefined indices. However, the presented methodology requires more data than is currently available in present-day reanalyses for subseasonal timescales. As such, the use of large climate model simulations for pretraining neural networks and subsequent transfer learning with present-day reanalyses to unlearn subseasonal model biases (or potentially longer timescales, such as seasonal-to-decadal) may not yield reliable results.

While we specifically explore a state-dependent shift in the input, other forms of biases or additional biases could require a different amount of (re)training data for a given prediction problem. We also note that the amount of data used for pre-training and validation could influence these results along with the architecture of the neural network and specific model hyperparameters chosen. Furthermore, the analysis and results presented are for a categorical application, whereas a regression problem may lead to different conclusions. Nevertheless, the perfect model framework used in this analysis is a viable starting point to explore whether this methodology is feasible for a specific application. Overall, using transfer learning to identify state-dependent biases relevant to predictability is feasible, but is currently not useful for this application on subseasonal timescales without more data from a longer observational record.

%

\acknowledgments
This material is based upon work supported by the U.S. Department of Energy (DOE), Office of Science, Office of Biological and Environmental Research (BER), Regional and Global Model Analysis (RGMA) component of the Earth and Environmental System Modeling Program under Award Number DE-SC0022070 and National Science Foundation (NSF) IA 1947282. This work was also supported by the U.S. National Science Foundation National Center for Atmospheric Research (NSF NCAR), which is a major facility sponsored by the NSF under Cooperative Agreement No. 1852977. The Energy Exascale Earth System Model (E3SM) project is funded by the U.S. Department of Energy, Office of Science, Office of Biological and Environmental Research (BER). E3SMv2 simulations were performed on a high-performance computing cluster provided by the BER Earth System Modeling program and operated by the Laboratory Computing Resource Center at Argonne National Laboratory. Data archiving of production simulations used resources of the National Energy Research Scientific Computing Center (NERSC), a DOE Office of Science User Facility supported by the Office of Science of the U.S. Department of Energy under Contract No. DE-AC02-05CH11231. Computing and data storage resources were provided by the Computational and Information Systems Laboratory (CISL) at NCAR. We thank all the scientists, software engineers, and administrators who contributed to the development of E3SMv2. MJM was also partly supported by a University of Maryland Grand Challenges Grants Program (GC17-2957817).


\datastatement
Native model output is available on NERSC at \url{https://portal.nersc.gov/archive/home/projects/e3sm/www/WaterCycle/E3SMv2/LR} with documentation located here: \url{https://e3sm-project.github.io/e3sm\_data\_docs}. Code to recreate this analysis can be found at \url{https://github.com/kjmayer/TL-XAI_E3SM}


\clearpage
\appendix

\appendixtitle{Supplemental Figures}

\begin{table}[t]
\caption{Table of the corresponding E3SMv2 ensemble member label for the members 1-21 mentioned in the main text.}
\begin{center}
\begin{tabular}{cccccccccccccccccccccc}
\hline
Member Number: & 1 & 2 & 3 & 4 & 5 & 6 & 7  \\
Corresponding E3SMv2 Member: & 101  & 111 & 121 & 131 & 141 & 151 & 161 \\
\hline

 & 8 & 9 & 10 & 11 & 12 & 13 & 14 \\
 & 171 & 181 & 191 & 201 & 211 & 221 & 231\\
\hline

 & 15 & 16 & 17 & 18 & 19 & 20 & 21 \\
 & 241 & 251 & 261 & 271 & 281 & 291 & 301\\
\hline

\end{tabular}
\end{center}
\end{table}

\begin{figure}[th]
  \noindent\includegraphics[width=39pc,angle=0]{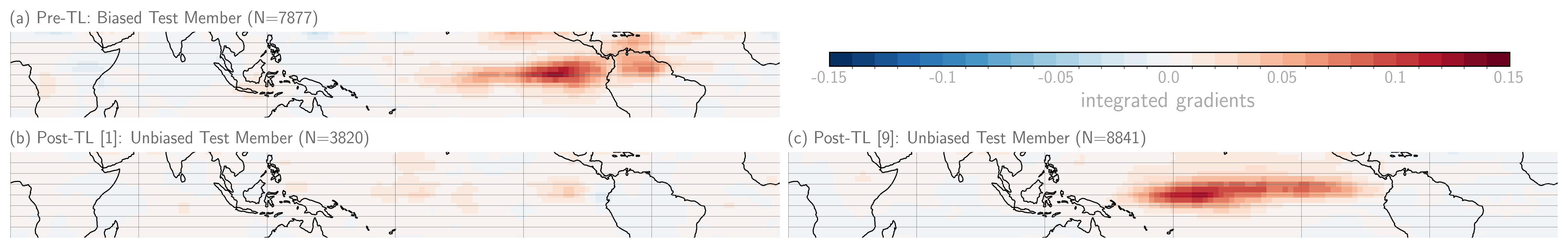}\\
  \caption{As in Figure \ref{XAI}, but for negative predictions from a convolutional neural network for (b) one or (c) nine retraining members.}\label{XAI_EXP3_CNN}
\end{figure}

\begin{figure}[th]
  \noindent\includegraphics[width=39pc,angle=0]{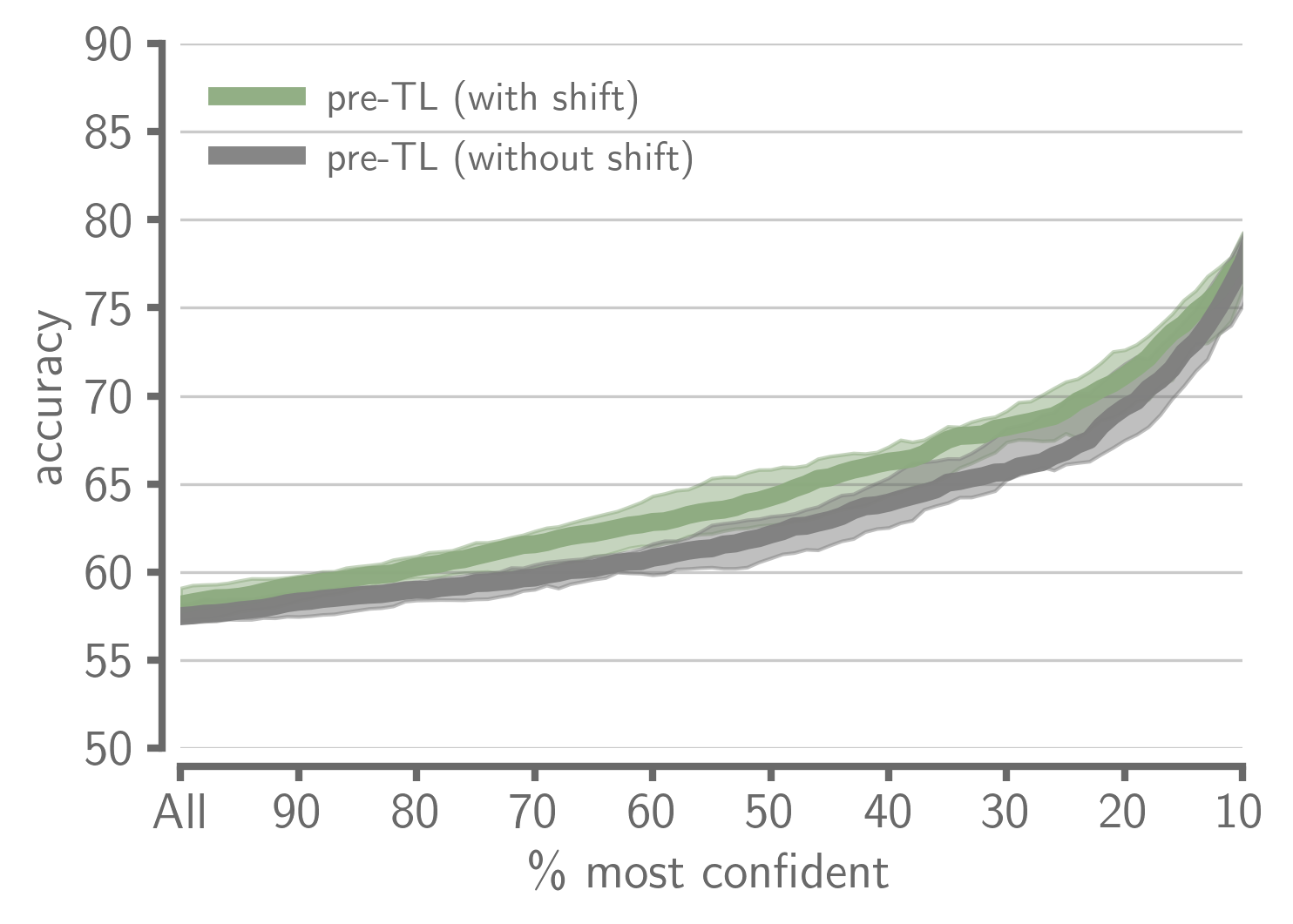}\\
  \caption{Confidence versus accuracy of the pre-transfer learning model on the test member with bias (green line) and without bias (grey line). Shading encompasses the skill across the ten randomly initialized neural networks.}\label{performance_ann1}
\end{figure}

\begin{figure}[th]
  \noindent\includegraphics[width=39pc,angle=0]{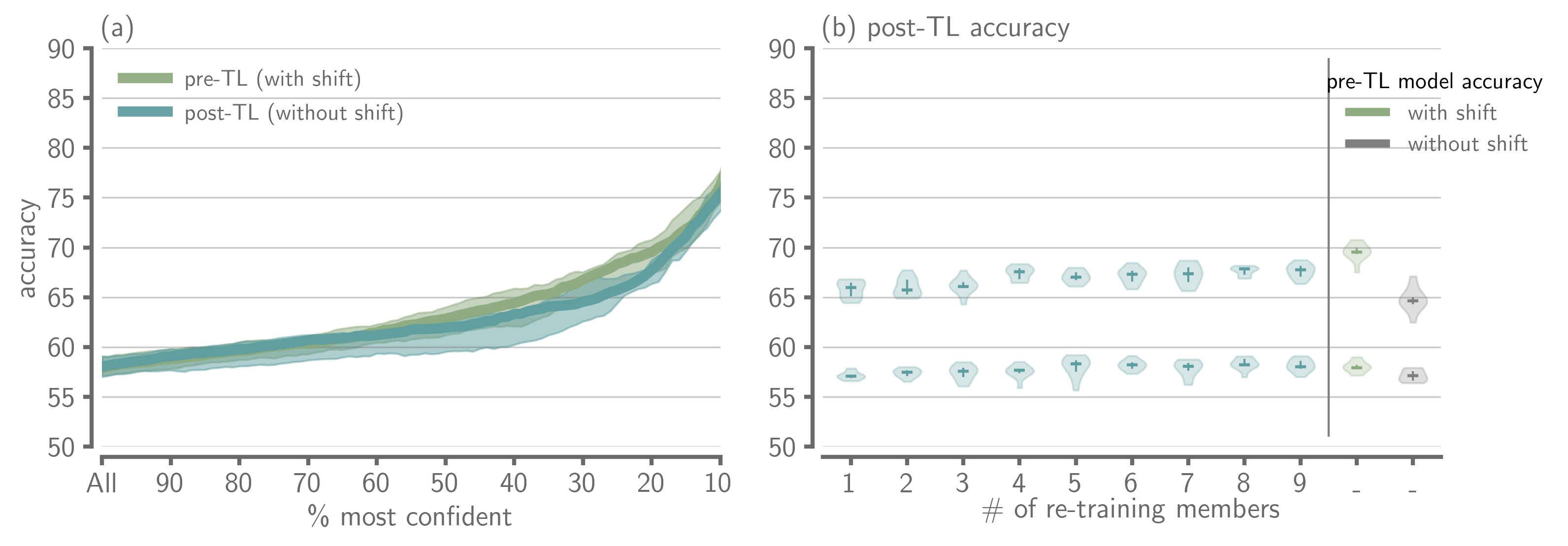}\\
  \caption{As in Figure \ref{performance}a and b, but for a testing member (member 1) that requires more than nine retraining members to achieve comparable performance to the pre-transfer learning network on the biased data.}\label{performance_EXP5}
\end{figure}

\begin{figure}[th]
  \noindent\includegraphics[width=39pc,angle=0]{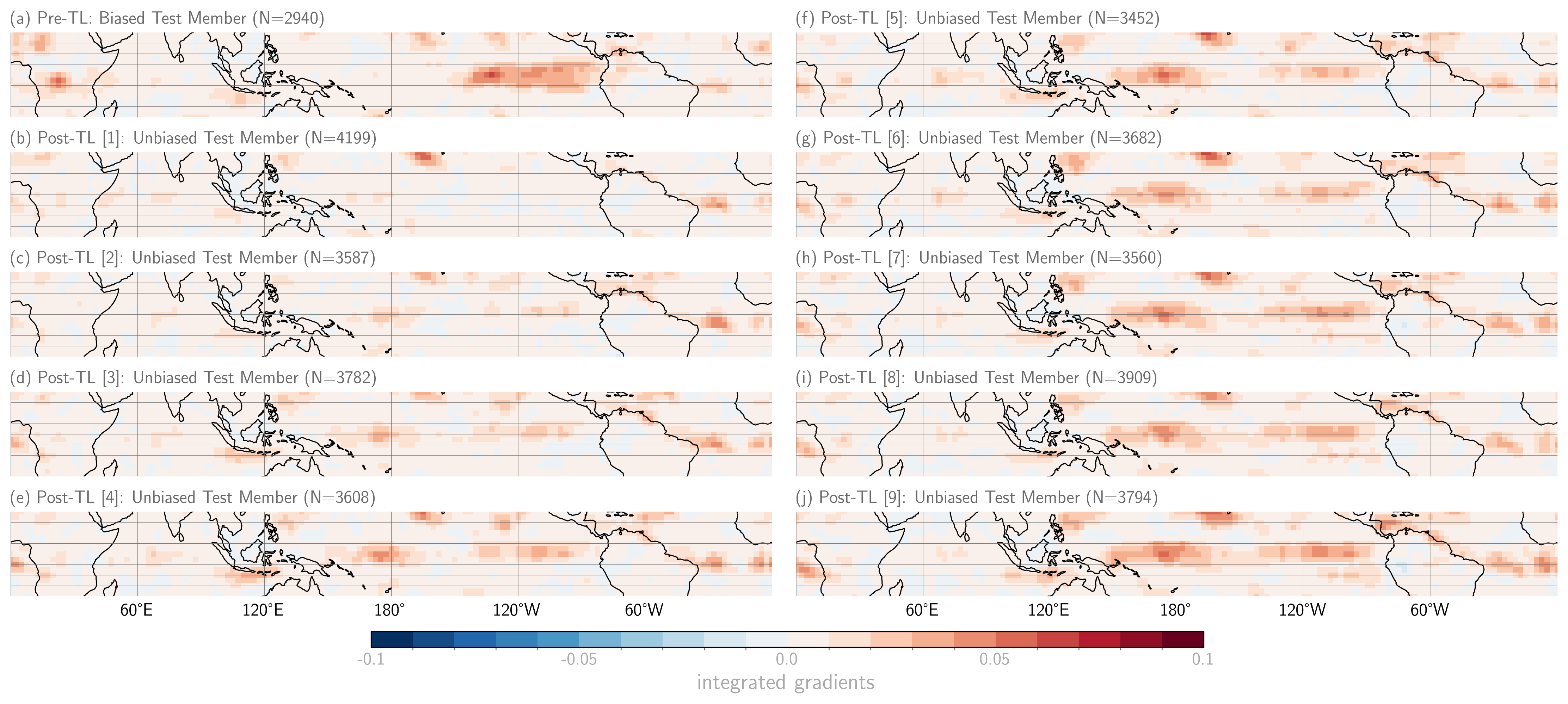}\\
  \caption{As in Figure \ref{XAI}, but for positive predictions.}\label{XAI_pos}
\end{figure}
\begin{figure}[th]
  \noindent\includegraphics[width=39pc,angle=0]{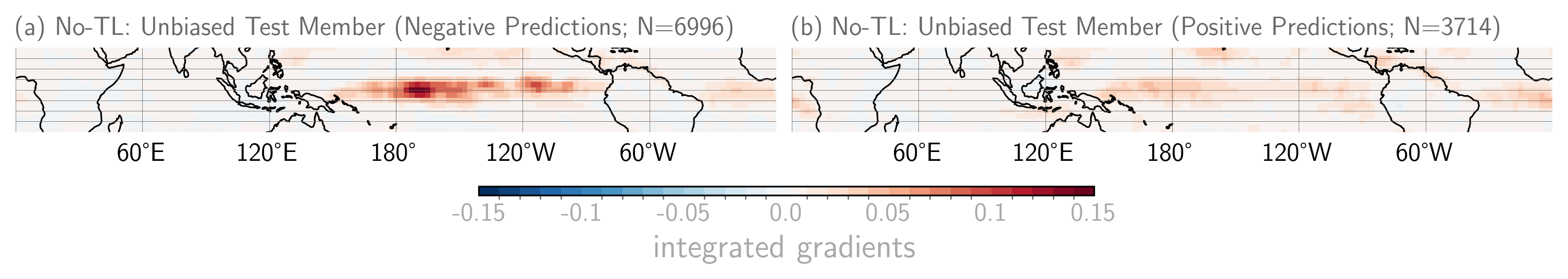}\\
  \caption{Integrated gradients result for confident and correct predictions when an artificial neural network with the same hyperparameters as the pre-transfer learning model is trained on unbiased training data (unbiased members 11-19, validation on unbiased member 20, and testing on unbiased member 21).}\label{noTLIG}
\end{figure}

\begin{figure}[th]
  \noindent\includegraphics[width=39pc,angle=0]{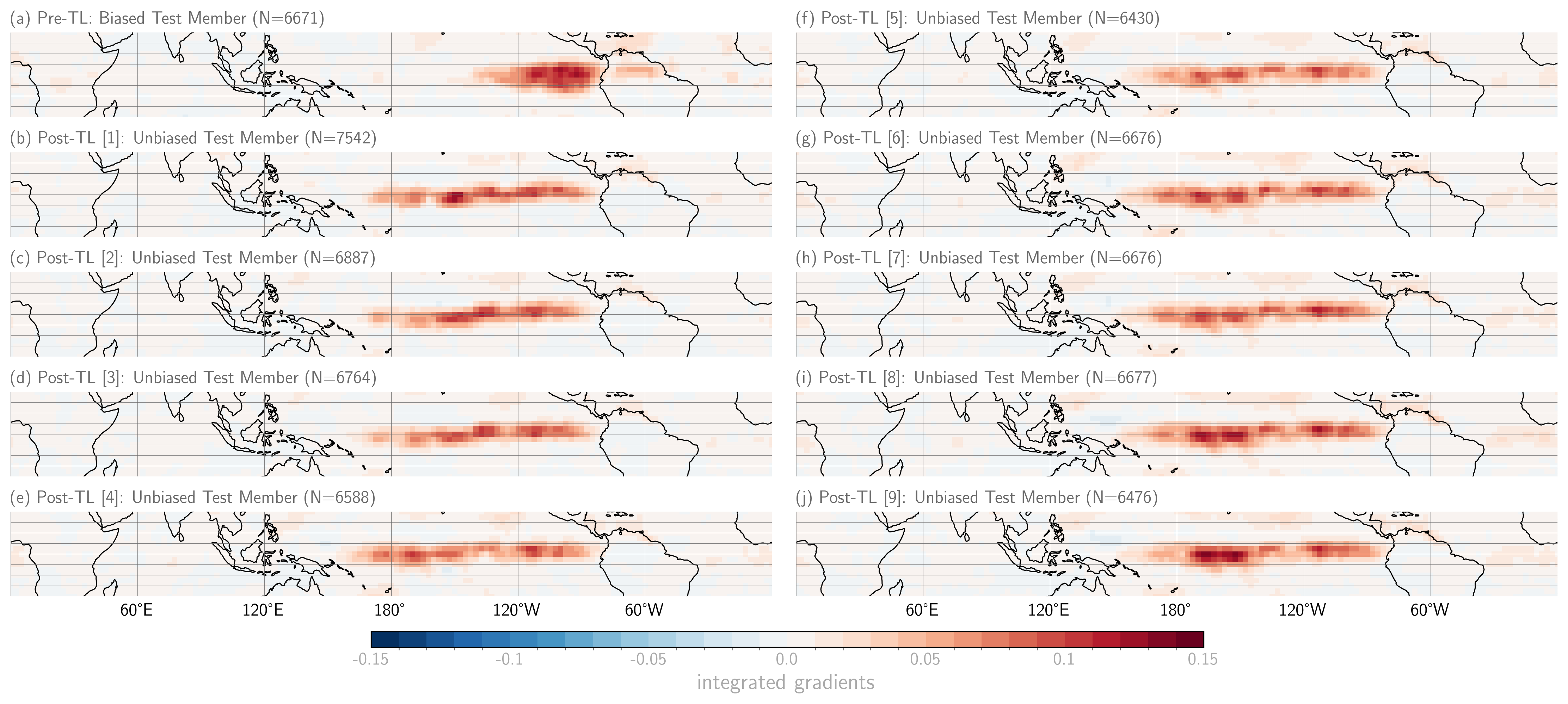}\\
  \caption{As in Figure \ref{XAI}, but for a testing member that requires more than nine retraining members (member 1).}\label{XAI_EXP5_neg}
\end{figure}

\begin{figure}[th]
  \noindent\includegraphics[width=39pc,angle=0]{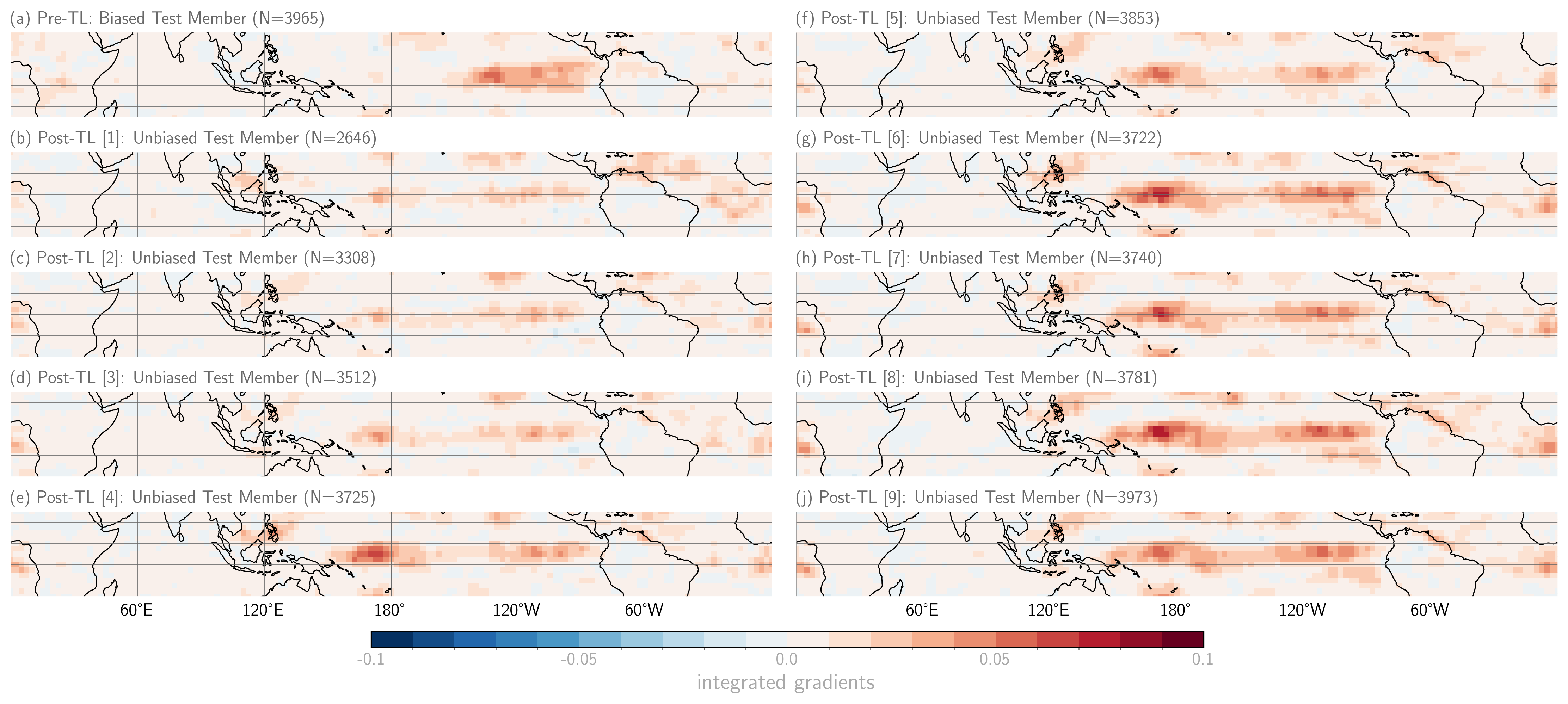}\\
  \caption{As in Figure \ref{XAI_EXP5_neg}, but for positive predictions.}\label{XAI_EXP5_pos}
\end{figure}

%



\clearpage
\bibliographystyle{ametsocV6}
\bibliography{references}

\begin{thebibliography}{35}
\providecommand{\natexlab}[1]{#1}
\providecommand{\url}[1]{\texttt{#1}}
\renewcommand{\UrlFont}{\rmfamily}
\providecommand{\urlprefix}{URL }
\expandafter\ifx\csname urlstyle\endcsname\relax
  \providecommand{\doi}[1]{https://doi.org/\discretionary{}{}{}#1}\else
  \providecommand{\doi}{https://doi.org/\discretionary{}{}{}\begingroup \urlstyle{rm}\Url}\fi
\providecommand{\eprint}[2][]{\url{#2}}

\bibitem[{Chapman et~al.(2021)Chapman, Subramanian, Xie, Sierks, Ralph,, and Kamae}]{chapman2021monthly}
Chapman, W.~E., A.~C. Subramanian, S.-P. Xie, M.~D. Sierks, F.~M. Ralph, and Y.~Kamae, 2021: {Monthly modulations of ENSO teleconnections: Implications for potential predictability in North America}. \textit{J.\ Climate}, \textbf{34~(14)}, 5899--5921, \doi{10.1175/JCLI-D-20-0391.1}.

\bibitem[{Davenport and Diffenbaugh(2021)Davenport, and Diffenbaugh}]{Davenport2021}
Davenport, F.~V., and N.~S. Diffenbaugh, 2021: Using machine learning to analyze physical causes of climate change: A case study of {U.S}. midwest extreme precipitation. \textit{Geophys. Res. Lett.}, \doi{10.1029/2021GL093787}.

\bibitem[{Golaz et~al.(2022)}]{Golaz2022}
Golaz, J.-C., and Coauthors, 2022: The {DOE} {E3SM} model version 2: Overview of the physical model and initial model evaluation. \textit{J. Adv. Model. Earth Syst.}, \textbf{14~(12)}, e2022MS003\,156.

\bibitem[{Gordon et~al.(2021)Gordon, Barnes,, and Hurrell}]{Gordon2021}
Gordon, E.~M., E.~A. Barnes, and J.~W. Hurrell, 2021: Oceanic harbingers of pacific decadal oscillation predictability in {CESM2} detected by neural networks. \textit{Geophys. Res. Lett.}, \textbf{48~(21)}.

\bibitem[{Ham et~al.(2021)Ham, Kim, Kim,, and On}]{Ham2021}
Ham, Y.-G., J.-H. Kim, E.-S. Kim, and K.-W. On, 2021: Unified deep learning model for el ni{\~n}o/southern oscillation forecasts by incorporating seasonality in climate data. \textit{Sci Bull. Fac. Agric. Kyushu Univ.}, \textbf{66~(13)}, 1358--1366.

\bibitem[{Ham et~al.(2019)Ham, Kim,, and Luo}]{Ham2019}
Ham, Y.-G., J.-H. Kim, and J.-J. Luo, 2019: Deep learning for multi-year {ENSO} forecasts. \textit{Nature}, \textbf{573~(7775)}, 568--572, \doi{10.1038/s41586-019-1559-7}.

\bibitem[{Henderson et~al.(2016)Henderson, Maloney,, and Barnes}]{henderson2016influence}
Henderson, S.~A., E.~D. Maloney, and E.~A. Barnes, 2016: The influence of the madden--julian oscillation on northern hemisphere winter blocking. \textit{Journal of Climate}, \textbf{29~(12)}, 4597--4616.

\bibitem[{Hoskins and Ambrizzi(1993)Hoskins, and Ambrizzi}]{Hoskins1993}
Hoskins, B.~J., and T.~Ambrizzi, 1993: Rossby wave propagation on a realistic longitudinally varying flow. \textit{J. Atmos. Sci.}, \textbf{50~(12)}, 1661--1671.

\bibitem[{Johnson et~al.(2014)Johnson, Collins, Feldstein, L'Heureux,, and Riddle}]{Johnson2014}
Johnson, N.~C., D.~C. Collins, S.~B. Feldstein, M.~L. L'Heureux, and E.~E. Riddle, 2014: Skillful wintertime north american temperature forecasts out to 4 weeks based on the state of {ENSO} and the {MJO}. \textit{Weather Forecast.}, \textbf{29~(1)}, 23--38.

\bibitem[{Kadow et~al.(2020)Kadow, Hall,, and Ulbrich}]{Kadow2020}
Kadow, C., D.~M. Hall, and U.~Ulbrich, 2020: Artificial intelligence reconstructs missing climate information. \textit{Nat. Geosci.}, \textbf{13~(6)}, 408--413.

\bibitem[{Kang et~al.(2020)Kang, Kim, Ahn, Neale, Lee,, and Gleckler}]{kang2020role}
Kang, D., D.~Kim, M.-S. Ahn, R.~Neale, J.~Lee, and P.~J. Gleckler, 2020: {The role of the mean state on MJO simulation in CESM2 ensemble simulation}. \textit{Geophys.\ Res.\ Lett.}, \textbf{47~(24)}, e2020GL089\,824, \doi{10.1029/2020GL089824}.

\bibitem[{Kingma and Ba(2014)Kingma, and Ba}]{kingma2014adam}
Kingma, D.~P., and J.~Ba, 2014: Adam: A method for stochastic optimization. \textit{arXiv preprint arXiv:1412.6980}.

\bibitem[{Kumar and Hoerling(1998)Kumar, and Hoerling}]{kumar1998annual}
Kumar, A., and M.~P. Hoerling, 1998: Annual cycle of pacific--north american seasonal predictability associated with different phases of enso. \textit{Journal of Climate}, \textbf{11~(12)}, 3295--3308.

\bibitem[{Madden and Julian(1971)Madden, and Julian}]{Madden1971}
Madden, R.~A., and P.~R. Julian, 1971: Detection of a 40--50 day oscillation in the zonal wind in the tropical pacific. \textit{J. Atmos. Sci.}, \textbf{28~(5)}, 702--708.

\bibitem[{Madden and Julian(1972)Madden, and Julian}]{Madden1972}
Madden, R.~A., and P.~R. Julian, 1972: Description of {Global-Scale} circulation cells in the tropics with a 40--50 day period. \textit{J. Atmos. Sci.}, \textbf{29~(6)}, 1109--1123.

\bibitem[{Madden and Julian(1994)Madden, and Julian}]{Madden1994}
Madden, R.~A., and P.~R. Julian, 1994: Observations of the 40--50-day tropical {Oscillation---A} review. \textit{Mon. Weather Rev.}, \textbf{122~(5)}, 814--837.

\bibitem[{Mamalakis et~al.(2022)Mamalakis, Ebert-Uphoff,, and Barnes}]{Mamalakis2022}
Mamalakis, A., I.~Ebert-Uphoff, and E.~A. Barnes, 2022: Neural network attribution methods for problems in geoscience: A novel synthetic benchmark dataset. \textit{Environmental Data Science}, \textbf{1}, e8.

\bibitem[{Mariotti et~al.(2020)}]{Mariotti2020}
Mariotti, A., and Coauthors, 2020: Windows of opportunity for skillful forecasts subseasonal to seasonal and beyond. \textit{Bull. Am. Meteorol. Soc.}

\bibitem[{Mayer and Barnes(2021)Mayer, and Barnes}]{Mayer2021}
Mayer, K.~J., and E.~A. Barnes, 2021: Subseasonal forecasts of opportunity identified by an explainable neural network. \textit{Geophys. Res. Lett.}

\bibitem[{Mayer and Barnes(2022)Mayer, and Barnes}]{Mayer2022}
Mayer, K.~J., and E.~A. Barnes, 2022: Quantifying the effect of climate change on midlatitude subseasonal prediction skill provided by the tropics. \textit{Geophys. Res. Lett.}, \textbf{49~(14)}.

\bibitem[{Mayer et~al.(2024)Mayer, Chapman,, and Manriquez}]{Mayer2024}
Mayer, K.~J., W.~E. Chapman, and W.~A. Manriquez, 2024: Exploring the relative importance of the {MJO} and {ENSO} to north pacific subseasonal predictability. \textit{Geophys. Res. Lett.}, \textbf{51~(10)}.

\bibitem[{McGovern et~al.(2019)McGovern, Lagerquist, Gagne, Jergensen, Elmore, Homeyer,, and Smith}]{mcgovern2019making}
McGovern, A., R.~Lagerquist, D.~J. Gagne, G.~E. Jergensen, K.~L. Elmore, C.~R. Homeyer, and T.~Smith, 2019: Making the black box more transparent: Understanding the physical implications of machine learning. \textit{Bulletin of the American Meteorological Society}, \textbf{100~(11)}, 2175--2199.

\bibitem[{Molina et~al.(2021)Molina, Gagne,, and Prein}]{Molina2021}
Molina, M.~J., D.~J. Gagne, and A.~F. Prein, 2021: A benchmark to test generalization capabilities of deep learning methods to classify severe convective storms in a changing climate. \textit{Earth Space Sci.}, \textbf{8~(9)}.

\bibitem[{Mundhenk et~al.(2018)Mundhenk, Barnes, Maloney,, and Baggett}]{Mundhenk2018}
Mundhenk, B.~D., E.~A. Barnes, E.~D. Maloney, and C.~F. Baggett, 2018: Skillful empirical subseasonal prediction of landfalling atmospheric river activity using the {Madden–Julian} oscillation and quasi-biennial oscillation. \textit{npj Climate and Atmospheric Science}, \textbf{1~(1)}, 20\,177.

\bibitem[{Philander(1985)}]{philander1985nino}
Philander, S., 1985: El ni{\~n}o and la ni{\~n}a. \textit{Journal of Atmospheric Sciences}, \textbf{42~(23)}, 2652--2662.

\bibitem[{Phillips et~al.(2020)Phillips, C., Fasullo, P.,, and Simpson}]{Phillips2020}
Phillips, A.~S., D.~C., S.~Fasullo, J., D.~P., and I.~R. Simpson, 2020: Assessing climate variability and change in model large ensembles: A user's guide to the ``climate variability diagnostics package for large ensembles.

\bibitem[{Rader et~al.(2022)Rader, Barnes, Ebert-Uphoff,, and Anderson}]{Rader2022}
Rader, J.~K., E.~A. Barnes, I.~Ebert-Uphoff, and C.~Anderson, 2022: Detection of forced change within combined climate fields using explainable neural networks. \textit{J. Adv. Model. Earth Syst.}, \textbf{14~(7)}.

\bibitem[{Sundararajan et~al.(2017)Sundararajan, Taly,, and Yan}]{Sundararajan2017}
Sundararajan, M., A.~Taly, and Q.~Yan, 2017: Axiomatic attribution for deep networks. \eprint{1703.01365}.

\bibitem[{Tan et~al.(2018)Tan, Sun, Kong, Zhang, Yang,, and Liu}]{Tan2018}
Tan, C., F.~Sun, T.~Kong, W.~Zhang, C.~Yang, and C.~Liu, 2018: A survey on deep transfer learning. \textit{Artificial Neural Networks and Machine Learning -- {ICANN} 2018}, Springer International Publishing, 270--279.

\bibitem[{Trenberth(1997)}]{Trenberth1997}
Trenberth, K.~E., 1997: The definition of el nino. \textit{Bull. Am. Meteorol. Soc.}, \textbf{78~(12)}, 2771--2778.

\bibitem[{Tseng et~al.(2018)Tseng, Barnes,, and Maloney}]{Tseng2018}
Tseng, K.-C., E.~A. Barnes, and E.~D. Maloney, 2018: Prediction of the midlatitude response to strong {Madden-Julian} oscillation events on {S2S} time scales: {PREDICTION} {OF} {Z500} {AT} {S2S} {TIME} {SCALES}. \textit{Geophys. Res. Lett.}, \textbf{45~(1)}, 463--470.

\bibitem[{Wang and Robertson(2019)Wang, and Robertson}]{Wang2019}
Wang, L., and A.~W. Robertson, 2019: Week 3--4 predictability over the united states assessed from two operational ensemble prediction systems. \textit{Clim. Dyn.}, \textbf{52~(9)}, 5861--5875.

\bibitem[{Wei et~al.(2021)Wei, Subramanian, Karnauskas, DeMott, Mazloff,, and Balmaseda}]{wei2021tropical}
Wei, H.-H., A.~C. Subramanian, K.~B. Karnauskas, C.~A. DeMott, M.~R. Mazloff, and M.~A. Balmaseda, 2021: {Tropical Pacific Air-Sea Interaction Processes and Biases in CESM2 and Their Relation to El Ni{\~n}o Development}. \textit{Journal of Geophysical Research: Oceans}, \textbf{126~(6)}, e2020JC016\,967, \doi{10.1029/2020JC016967}.

\bibitem[{White et~al.(2017)}]{CWhite2017}
White, C.~J., and Coauthors, 2017: Potential applications of subseasonal-to-seasonal ({S2S}) predictions. \textit{Met. Apps}, \textbf{24~(3)}, 315--325.

\bibitem[{White et~al.(2021)}]{CWhite2021}
White, C.~J., and Coauthors, 2021: Advances in the application and utility of subseasonal-to-seasonal predictions. \textit{Bull. Am. Meteorol. Soc.}, \textbf{-1~(aop)}, 1--57.

\end{thebibliography}

\end{document}